\documentclass[dvipdfmx]{PoS}
\usepackage{amsmath}
\title{On the condition for correct convergence
in the complex Langevin method\thanks{KEK-TH/1946}}

\ShortTitle{On the condition for correct convergence in the complex Langevin method}

\author{\speaker{Shinji Shimasaki}\\
        Research and Education Center for Natural Sciences, Keio University, 4-1-1 Hiyoshi, \\
        Yokohama, Kanagawa 223-8521, Japan\\
        E-mail: \email{shinji.shimasaki@keio.jp}}

\author{Keitaro Nagata\\
        KEK , High Energy Accelerator Research Organization, 1-1 Oho, \\
        Tsukuba, Ibaraki 305-0801, Japan\\
        E-mail: \email{knagata@post.kek.jp}}

\author{Jun Nishimura\\
        KEK , High Energy Accelerator Research Organization, 1-1 Oho, \\
        Tsukuba, Ibaraki 305-0801, Japan\\
        Graduate University for Advanced Studies (SOKENDAI), 1-1 Oho, \\
        Tsukuba, Ibaraki 305-0801, Japan\\
        E-mail: \email{jnishi@post.kek.jp}}

\abstract{
The complex Langevin method (CLM) provides a promising way 
to perform the path integral with a complex action 
using a stochastic equation for complexified dynamical variables. 
It is known, however, that the method gives wrong results in some cases, 
while it works, for instance, 
in finite density QCD in the deconfinement phase or in the heavy dense limit. 
Here we revisit the argument for justification of the CLM
and point out a subtlety in using the time-evolved observables, 
which play a crucial role in the argument. 
This subtlety requires that the probability distribution 
of the drift term should fall off exponentially or faster 
at large magnitude.
We demonstrate our claim in some examples such as
chiral Random Matrix Theory and show that our criterion is 
indeed useful in judging whether the results obtained by the CLM 
are trustable or not.
}

\FullConference{34th annual International Symposium on Lattice Field Theory\\
		24-30 July 2016\\
		University of Southampton, UK}

\begin{document}

\section{Introduction}

The complex Langevin method (CLM) \cite{Parisi:1984cs,Klauder:1983sp}
is a promising approach to the complex action problem based
on a stochastic process for complexified variables.
Although it is well-known that the method does not always 
give correct results, 
the range of applicability has been substantially enlarged
thanks to the recent development of a new technique 
such as gauge cooling \cite{Seiler:2012wz}.
In particular, the gauge cooling has enabled
the application of the method to finite density QCD
in the deconfined phase \cite{Sexty:2013ica,Fodor:2015doa}.

The argument for justification of the CLM was given
in ref.~\cite{Aarts:2009uq,Aarts:2011ax}.
There, a crucial step was to shift
the time evolution of the probability distribution 
of the complexified variables
to that of observables.
This is possible only if the integration by parts used there
is valid.
For this reason,
the CLM fails
when the probability distribution does not fall off fast enough 
in the asymptotic region \cite{Aarts:2009uq,Aarts:2011ax} 
(the excursion problem)
or in the region near singularities of the drift term when
they exist \cite{Nishimura:2015pba} (the singular drift problem).
It was shown that 
not only the excursion problem \cite{Seiler:2012wz} but also the 
singular drift problem \cite{Nagata:2016alq} may be cured by 
the gauge cooling.


Here we revisit the argument 
for justification of the CLM
and point out a subtlety in the use of 
time-evolved observables \cite{Nagata:2016vkn}.
In the previous argument, it was implicitly assumed that 
the time-evolved observables can be used for an infinitely long time.
We point out that this is a too strong assumption,
which is not necessarily satisfied even in cases where 
the CLM gives correct results.
In fact, what is needed for justification is 
the use of the time-evolved observables for a finite but nonzero time.
This still requires that the probability distribution of 
the drift term should be suppressed
exponentially at large magnitude, which is slightly stronger
than the condition for the validity of the integration by parts.

The rest of this article is organized as follows. 
In section 2, we discuss the condition for 
justification of the CLM taking into account the subtlety
in the time-evolved observables.
In section 3, we demonstrate our condition in simple models.
Section 4 is devoted to a summary.

\section{New argument for justification of the CLM}

Let us consider a partition function
$Z = \int dx \, w(x)$
written in terms of a real variable $x$
with a complex weight $w(x)$.
In the CLM, we consider the Langevin equation for
the complexified variable $x\to z=x+iy$,
which takes the form
\begin{align}
z^{(\eta)} (t+\epsilon) 
= z^{(\eta)} (t) 
+ \epsilon \, v (z)
+ \sqrt{\epsilon} \, \eta(t) 
\label{dCLE}
\end{align}
in its discretized version.
Here, $v(z)$ represents the drift term,
which can be obtained by analytically continuing
$v(x)=w(x)^{-1}\partial w(x)/\partial x$,
and $\eta(t)$ is a real noise, which obeys
the probability distribution $\propto e^{-\frac{1}{4}\sum_t\eta^2(t)}$.
Below, we denote the expectation value with respect to $\eta$ 
as $\langle\cdots\rangle_\eta$.
Let us also extend the observable $\mathcal O(x)$ in the original model
to a holomorphic function $\mathcal O(z)$ of $z$ by analytic continuation
and define the expectation value of $\mathcal O(z)$ as
\begin{alignat}{3}
\Phi(t) = 
\langle  {\cal O}(x^{(\eta)} (t)+i y^{(\eta)} (t))
\rangle_{\eta} 
=
\int 
dx \, dy \, {\cal O}(x+iy) \, P(x,y;t) \ ,
\label{OP-rewriting}
\end{alignat}
where $P(x,y;t)$ is the probability distribution 
of $x^{(\eta)}(t)$ and $y^{(\eta)}(t)$ defined by 
$P(x,y;t) =\langle \delta (x - x^{(\eta)} (t) )
\, \delta (y - y^{(\eta)} (t) )
\rangle_\eta$.
The crucial issue for the CLM is
whether this quantity $\Phi(t)$, after taking the $t\rightarrow \infty$
and $\varepsilon\rightarrow 0$ limits, reproduces
the expectation value of $\mathcal O(x)$ with respect to 
the original path integral with a complex weight,
namely whether the identity
\begin{alignat}{3}
\lim_{t \rightarrow \infty} 
\lim_{\epsilon \rightarrow 0 } \, 
\Phi(t)
&= 
\frac{1}{Z} \int 
dx
\, {\cal O}(x) \, w(x)
\label{O-time-av-complex}
\end{alignat}
holds or not.

In order to address this issue, we 
consider the time-evolution of the expectation value $\Phi(t)$.
Using the Langevin equation (\ref{dCLE}), we can write it as
\begin{alignat}{3}
\Phi(t + \epsilon)  &= 
\int 
dx \, dy \, {\cal O}_{\epsilon}(x+iy) \, 
P(x,y;t) \ ,
\label{OP-rewriting-P2} \\
{\cal O}_{\epsilon}(z) 
&=   \frac{1}{ {\cal N}}
\int d\eta \, 
e^{-\frac{1}{4}  \eta^{2} } 
{\cal O} \Big(z+\epsilon \, v(z)+\sqrt{\epsilon}\, \eta  \Big) \ ,
\label{OP-rewriting-P3}
\end{alignat}
where $\mathcal N$ is a normalization constant.
Since $\mathcal O(z)$ and $v(z)$ are holomorphic,
so is $\mathcal O_\epsilon(z)$.
By ``holomorphic'', we actually mean that the functions
are holomorphic in the region visited by the Langevin process.
In particular, we allow the case in which the functions have
singularities, which are measure zero in the configuration space.
Expanding (\ref{OP-rewriting-P3}) with respect to $\epsilon$ and
using the holomorphy of $\mathcal{O}(z)$, 
we can rewrite (\ref{OP-rewriting-P2}) as
\begin{alignat}{3}
\Phi(t + \epsilon)  
&= 
\sum_{n=0}^{\infty}
  \frac{1}{n!} \, \epsilon^n  
\int 
dx \, dy \, 
\Big( \mbox{\bf :} \tilde{L}^n  \mbox{\bf :} \,  {\cal O}(z) \Big) 
P(x,y;t) \ ,
\label{OP-rewriting-P3b}
\end{alignat}
where 
$\tilde{L}
= \left( \partial/\partial z + v(z)
\right)
\partial/\partial z$ 
and the symbol $\mbox{\bf :} \cdots \mbox{\bf :}$ implies
that the derivative operators are moved to the right as in
$\mbox{\bf :} ( f(x) + \partial)^2 \mbox{\bf :}
= f(x)^2 + 2f(x)\partial + \partial^2$.

If the $\epsilon$-expansion (\ref{OP-rewriting-P3b})
is valid, 
one can take the $\epsilon \to 0$ limit and get
\begin{alignat}{3}
\frac{d}{dt} \, \Phi(t)
&=  \int dx \, dy \, 
\Big\{ \tilde{L} \, {\cal O}(z) \Big\}
\, P(x,y;t)  \ .
\label{OP-rewriting-P3b-cont-lim}
\end{alignat}
In fact, it is known 
from the previous
argument \cite{Aarts:2009uq,Aarts:2011ax}
that (\ref{OP-rewriting-P3b-cont-lim}) does not hold
when the integration by parts, which is necessary in showing it,
becomes invalid.
In the present argument, on the other hand,
the possible failure of (\ref{OP-rewriting-P3b-cont-lim}) should 
be attributed to the breakdown of the $\epsilon$-expansion 
(\ref{OP-rewriting-P3b}).
Indeed, $\tilde L^n$ involves the $n$-th power of the drift term, 
which may become infinitely large. 
Therefore, the integral in (\ref{OP-rewriting-P3b}) can be 
divergent for large enough $n$.
What we have done so far is just an alternative presentation 
of the known problem that (\ref{OP-rewriting-P3b-cont-lim}) can be violated.
However, we will see below that
a similar argument for a finite time-evolution of $\Phi(t)$
gives rise to a condition, which is stronger than the one needed
for the validity of (\ref{OP-rewriting-P3b-cont-lim}).

The time-evolution of the expectation value $\Phi(t)$ 
for a finite $\tau$ can be obtained formally by
repeating the above argument for $\tilde L^n \mathcal O(z)$ as
\begin{alignat}{3}
 \Phi(t+\tau)
&=  \sum_{n=0}^{\infty}
  \frac{1}{n!} \, \tau^n
\int dx \, dy \, 
\Big\{  \tilde{L}^n \, {\cal O}(z) \Big\}
\, P(x,y;t)  \ .
\label{OP-rewriting-P3b-cont-lim-exp}
\end{alignat}
In order for this expression to be valid, however,
it is not sufficient to require that the integral 
appearing in the infinite series is convergent.
What also matters is the convergence radius of the infinite series.
In the previous argument, (\ref{O-time-av-complex}) was proved 
by assuming implicitly that the convergence radius is infinite.
Actually, this assumption is too strong and can be relaxed
if we employ the induction with respect 
to the Langevin time \cite{Nagata:2016vkn}.
What is needed to prove (\ref{O-time-av-complex}) then is 
that the convergence radius, which depends on $t$ in general, 
is bounded from below as a function of $t$.

In what follows, we discuss the explicit condition 
for the expression (\ref{OP-rewriting-P3b-cont-lim-exp}) 
to be valid \cite{Nagata:2016vkn}.
Let us define the probability distribution of the magnitude 
of the drift $u(z)=|v(z)|$ by
\begin{alignat}{3}
p(u;t) \equiv  \int dx \, dy \, \delta(u(z)-u) \, P(x,y;t) \ .
\label{def-u-prob}
\end{alignat}
Then, the most dominant contribution for each $n$ 
in (\ref{OP-rewriting-P3b})
and (\ref{OP-rewriting-P3b-cont-lim-exp}) may be written as
\begin{alignat}{3}
\int dx \, dy \, u(z)^n \, P(x,y;t) 
= \int_0^\infty  du \, u^n \, p(u;t) \ .
\label{simplified-integral}
\end{alignat}
In order for this to be finite for arbitrary $n$, 
$p(u;t)$ should fall off at large $u$ faster than
any power law. This condition is required
for the $\epsilon$-expansion 
(\ref{OP-rewriting-P3b})
and the $\tau$-expansion (\ref{OP-rewriting-P3b-cont-lim-exp})
to be valid.
In order for the 
$\tau$-expansion (\ref{OP-rewriting-P3b-cont-lim-exp})
to be valid for a finite $\tau$,
we need to require further that 
the convergence radius of the infinite series should be non-zero.
For instance, in cases where the probability distribution
of the drift term
is suppressed exponentially as $p(u;t)\sim e^{-\kappa u}$ 
for some $\kappa>0$, 
the convergence radius is estimated
as $\tau\sim \kappa$.
This implies that, 
in order for the $\tau$-expansion (\ref{OP-rewriting-P3b-cont-lim-exp}) 
to have a nonzero convergence radius, 
$p(u; t)$ has to fall off exponentially or faster.
Note that this condition is slightly stronger than 
the one obtained from the validity of the $\epsilon$-expansion,
which is equivalent to the validity of the integration by parts 
discussed in refs.~\cite{Aarts:2009uq,Aarts:2011ax}.
Therefore, our condition
may be viewed as a necessary and sufficient condition 
for justification of the CLM.

While our argument above is given in a single variable case for simplicity,
we can extend it to more general cases 
with multiple variables including the lattice gauge theory. 
We can also include the gauge cooling in the argument
as in ref.~\cite{Nagata:2015uga}.
For a comprehensive presentation of the argument including 
such generalizations, see ref.~\cite{Nagata:2016vkn}.

\section{Demonstration of our condition}

In this section, we demonstrate our new condition 
for justification of the CLM in some models.
First, we discuss two one-variable models, 
in which the CLM fails in some parameter region 
due to the singular drift problem or
the excursion problem \cite{Nagata:2016vkn}.
According to our argument above, these failures should be
attributed to the appearance of 
a large drift and hence they are understood in a unified manner.
Next, we discuss the chiral Random Matrix Theory (cRMT) \cite{Nagata},
which suffers from the singular 
drift problem \cite{Mollgaard:2013qra,Nagata:2016alq}
at small quark mass.
This example clearly shows that our condition is valid also
in a multi-variable case.

\subsection{A model with a singular drift}

\begin{figure}[t]
\centering
\includegraphics[width=5cm]{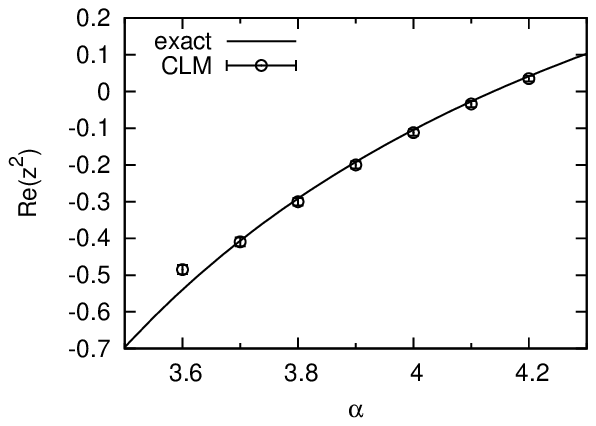}\hspace{10mm}
\includegraphics[width=5cm]{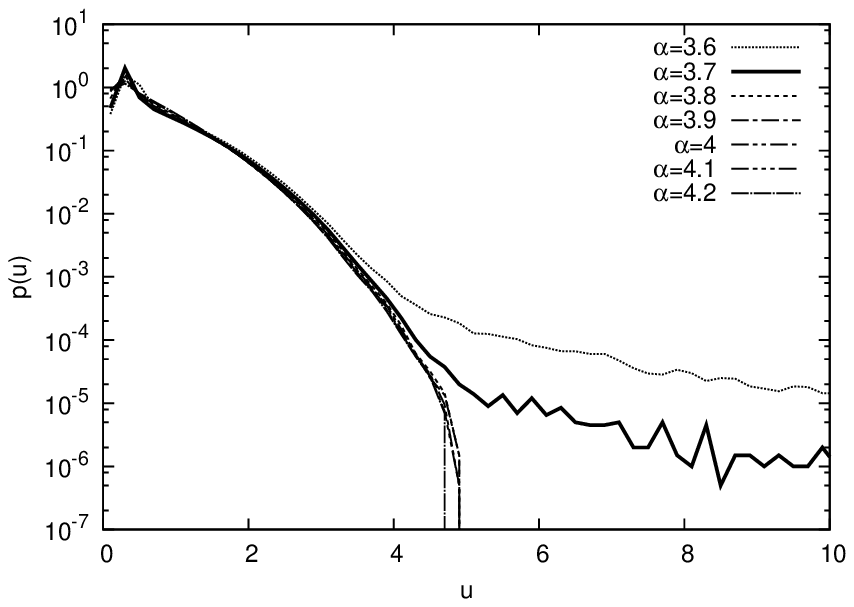}
\caption{(Left) The real part of the expectation value of ${\cal O}(z)=z^2$
obtained by the CLM is plotted against $\alpha$ for $p=4$.
The solid line represents the exact result.
(Right) The probability distribution $p(u)$ 
of the magnitude $u=|v|$ of the drift term 
is shown for various $\alpha$ within $3.6 \le \alpha \le 4.2$
in a semi-log plot.}
\label{singular_drift_result_x2}
\end{figure}

Here we consider a model with a singular drift term, 
whose partition function is given by
\begin{eqnarray}
Z = \int dx \, w(x) \ ,\quad
w(x) = (x+i\alpha)^p \, e^{-x^2/2} \ ,
\label{part-1var}
\end{eqnarray}
where $x$ is a real variable and $\alpha$ and $p$ are real parameters.
We perform simulations for $p=4$ and various $\alpha$ 
with the step-size $\epsilon=10^{-5}$.
The initial configuration is taken to be $z=0$ 
and the first $3\times 10^5$ steps are discarded 
for thermalization. 
After thermalization, we make $10^{10}$ steps and 
perform measurements every $10^3$ steps.
In Fig.~\ref{singular_drift_result_x2} (Left), 
we plot the real part of the expectation value of $\mathcal O(z)=z^2$ 
against $\alpha$,
which shows that the CLM reproduces the correct results 
for $\alpha\gtrsim 3.7$.

According to our new argument, the CLM fails
when the probability distribution of the drift term
is not suppressed exponentially at large magnitude.
This is confirmed in Fig.~\ref{singular_drift_result_x2} (Right), 
in which 
we plot
the probability distribution of the magnitude of the drift term
for various $\alpha$.
We find that the fall-off of the distribution
is faster than exponential for $\alpha\ge 3.8$,
while it is a power law for $\alpha\le 3.7$ \cite{Nagata:2016vkn}.
The result for $\alpha=3.7$ seems to agree with the exact result
presumably because the discrepancy is too small to be measured.

\subsection{A model with a possibility of excursions}

\begin{figure}[t]
\centering
\includegraphics[width=5cm]{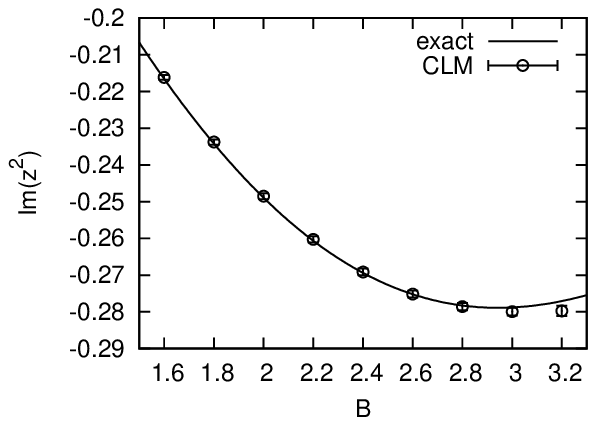}\hspace{10mm}
\includegraphics[width=5cm]{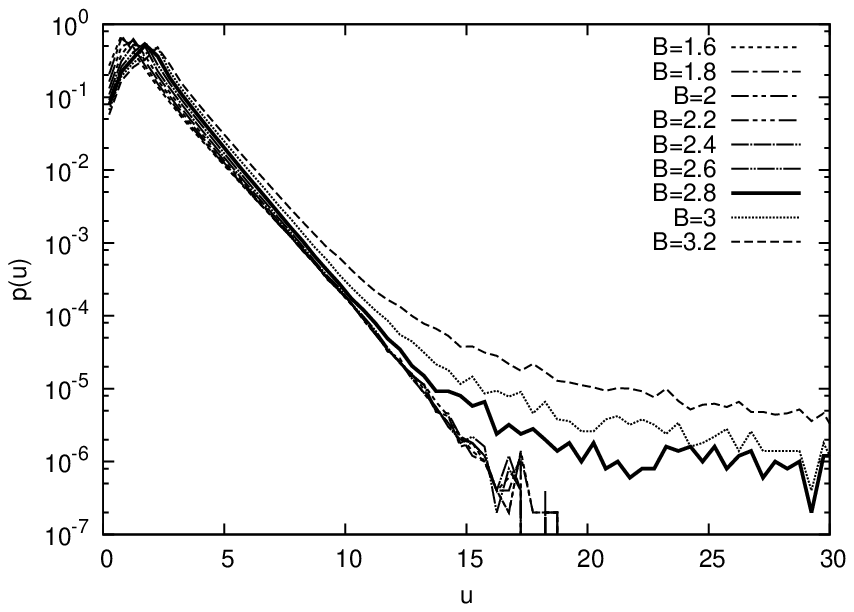}
\caption{(Left) The imaginary part of the expectation value 
of ${\cal O}(z)=z^2$ is plotted against $B$ for $A=1$.
The solid line represents the exact result. 
(Right) The probability distribution $p(u)$ 
for the magnitude $u=|v|$ of the drift term 
is shown for various $B$ within $1.6 \le B \le 3.2$
in a semi-log plot.
}
\label{skirt_result_x2}
\end{figure}

Here we consider a model with the excursion problem, 
whose partition function is \cite{Aarts:2013uza}
\begin{align}
Z = \int dx \, w(x) \ ,\quad
w(x) = e^{-\frac{1}{2}(A+iB)x^2-\frac{1}{4}x^4} \ ,
\label{part-1var-2}
\end{align}
where $x$ is a real variable and $A$ and $B$ are real parameters.
The simulation parameters are the same as those in the previous example
except for the use of an adaptive step-size with
the step-size $\epsilon$ being replaced by $\epsilon=0.01/|v(z)|$ 
when the magnitude of the drift $|v(z)|$ exceeds $10^3$. 
This is needed to avoid the runaway problem, which occurs for $B\ge 3.0$.
In Fig.~\ref{skirt_result_x2} (Left), 
we plot the imaginary part of the expectation value 
of $\mathcal O(z)=z^2$ against $B$.
It shows that the CLM reproduces the correct results for $B\le 2.8$.
Our new argument is confirmed in Fig.~\ref{skirt_result_x2} (Right), 
in which the probability distribution of the magnitude of 
the drift term is plotted for various $B$.
We find that the distribution falls off exponentially for $B\le 2.6$, 
while it falls off by a power for $B\ge 2.8$ \cite{Nagata:2016vkn}.
The result at $B=2.8$ agrees with the exact result
presumably because the discrepancy is too small to be measured.

\subsection{chiral Random Matrix Theory}

\begin{figure}[t]
\centering
\includegraphics[width=4.5cm]{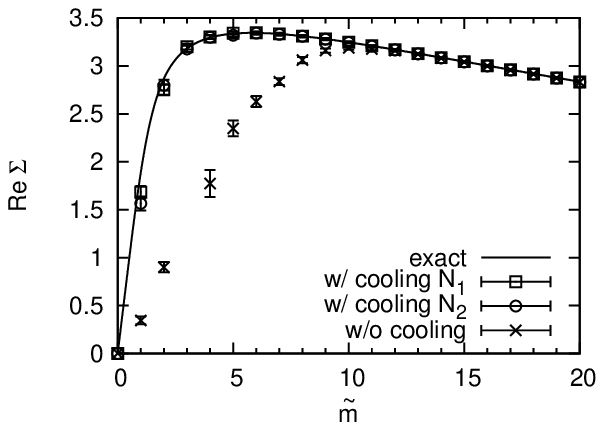}
\hspace{5mm}
\includegraphics[width=4.5cm]{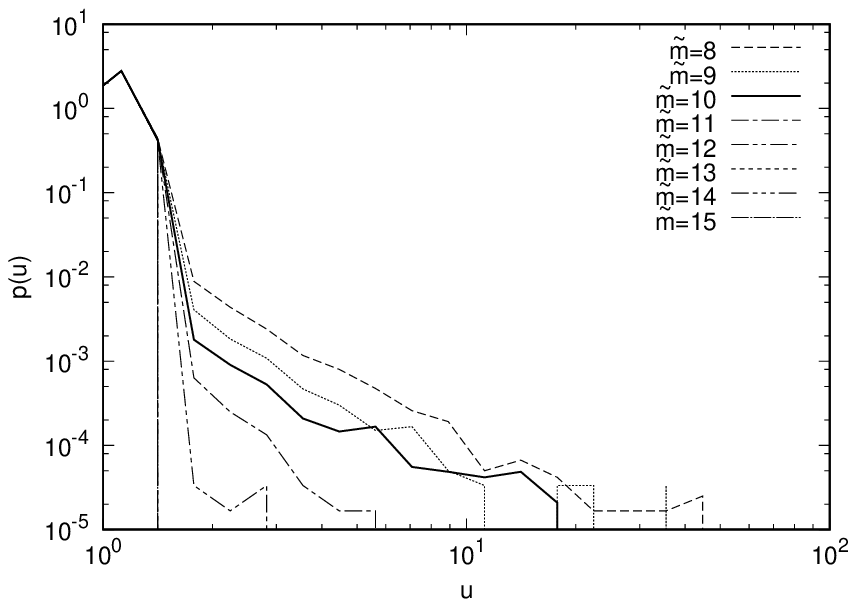}
\hspace{5mm}
\includegraphics[width=4.5cm]{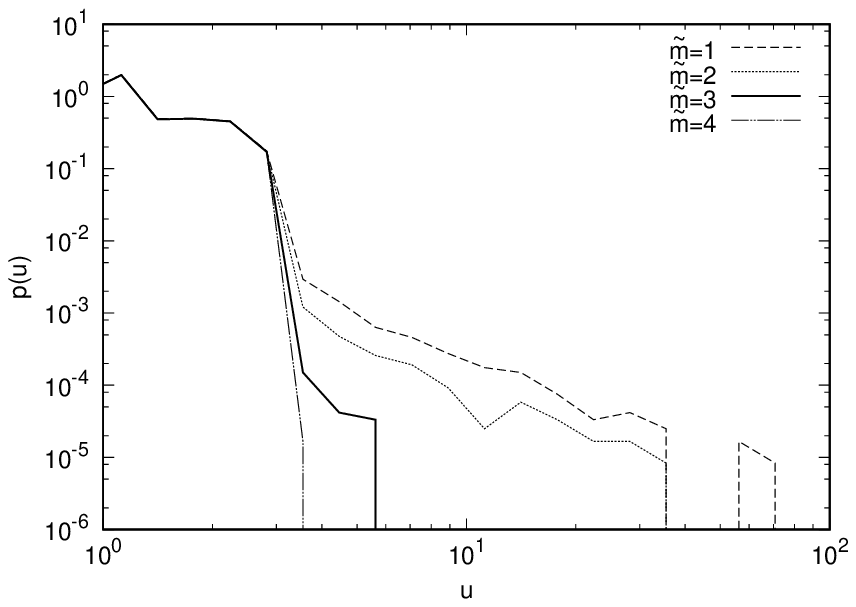}
\caption{
(Left) The real part of the expectation value of the chiral condensate 
is plotted against $\tilde m$.
The results are obtained by the CLM with or without gauge cooling. 
The solid line represents the exact result.
The probability distribution of the magnitude of the 
drift $p(u)$ (\protect\ref{p of crmt})
is shown in log-log plots for various $\tilde m$ in the cases 
without gauge cooling (Middle) and 
with the gauge cooling using the norm $N_1$ (Right).
}
\label{crmt fig}
\end{figure}

In order to demonstrate our condition in a multi-variable case,
we consider the cRMT for $N_f$ quarks with the degenerate mass $m$ 
and the chemical potential $\mu$.
The partition function is given by
\begin{align}
Z &= \int d\Phi_1d\Phi_2 \,  [\det (D+m)]^{N_{\rm f}} e^{-S_{\rm b}}  \ , 
\label{crmt}
\end{align}
where $\Phi_k$ $(k=1,2)$ are $N\times N$ general complex matrices.
The bosonic action $S_{\rm b}$ in (\ref{crmt}) is given by 
$S_{\rm b}=2N \sum_{k=1}^{2}{\rm Tr} (\Phi_k^\dagger \Phi_k)$ 
and the $2N\times 2N$ matrix $D$ is given as
\begin{align}
D&=\left( \begin{matrix}
0 & e^{\mu} \Phi_1 + e^{-\mu} \Phi_2 \\
-e^{-\mu} \Phi_1^\dagger - e^\mu \Phi_2^\dagger  & 0
\end{matrix} \right)  \ .
\end{align}
We apply the CLM for $N=30$, $N_{\rm f}=2$, $\tilde\mu\equiv\mu\sqrt{N}=2$ 
and various $\tilde m\equiv mN$.
In Fig.~\ref{crmt fig} (Left), 
we plot the real part of the chiral condensate 
$\Sigma = N^{-1} \partial \log Z/\partial m$
obtained with or without gauge cooling 
as a function of $\tilde m$ \cite{Nagata:2016alq}.
In this Figure, $N_1$ and $N_2$ stand for
the types of norm used for gauge cooling.
See ref.~\cite{Nagata:2016alq} for the details.
We find that the CLM reproduces the exact results 
for $\tilde m\gtrsim 10$ without gauge cooling
and for $\tilde m\gtrsim 1$ with gauge cooling using the norm $N_1$.

Next, we discuss the probability distribution of the magnitude 
of the drift term.
Let us denote the drift terms for $\Phi_i$ and $\Phi_i^\dagger$ 
by $F_i$ and $\bar F_i$ ($i=1,2$), respectively.
Then, the probability distribution may be defined as
\begin{align}
p(u)=\frac{1}{2N}
\sum_{i=1}^{2}
\sum_{a=1}^{N}\langle(\delta(u-v^{(a)}_i)+\delta(u-\bar v^{(a)}_i))\rangle \ ,
\label{p of crmt}
\end{align}
where $v_i^{(a)}$ and $\bar v_i^{(a)}$ ($a=1, \cdots , N$) 
are the eigenvalues of 
$(F_i^\dagger F_i)^{1/2}$ and $(\bar F_i^\dagger \bar F_i)^{1/2}$, 
respectively.
Note that this definition respects the $U(N)\times U(N)$ symmetry 
of the original cRMT.
In Fig.~\ref{crmt fig} (Middle) and (Right), 
we plot the probability distribution of the magnitude 
of the drift term (\ref{p of crmt}) 
against various $\tilde m$ in cases with or without gauge cooling.
We observe a power-law fall-off of the distribution
for $\tilde m\lesssim 12$ without gauge cooling 
and for $\tilde m\lesssim 3$ with gauge cooling using the norm $N_1$.
These regions agree with the regions where the CLM gives wrong results.

\section{Summary}

We revisited the argument for justification of 
the CLM, which was first given in ref.~\cite{Aarts:2009uq,Aarts:2011ax}.
In particular, we pointed out 
that the assumption made in the previous argument 
that time-evolved observables can be used
for an infinitely long time is too strong.
All we need to show (\ref{O-time-av-complex})
is the use of time-evolved observables for a finite but nonzero time. 
This still requires that the probability distribution of the drift term 
should fall off exponentially or faster at large magnitude. 
Our new condition can be used to probe the two possible problems in the CLM,
namely the excursion problem and the singular drift problem,
in a unified manner and to judge whether the results obtained 
by the CLM are trustable or not.
This was demonstrated
in two one-variable models and the cRMT,
where it was shown that the CLM reproduces the exact results 
when the probability distribution of 
the drift term falls off exponentially or faster 
at large magnitude.
Obviously, our condition should be of particular use
in applying the CLM to cases in which the results are not known
a priori. See ref.~\cite{Nagata2} for 
an application to finite density QCD.

\section*{Acknowledgments}
This work was supported by Grant-in-Aid for Scientific Research 
(No.~26800154 for K.N.\ and No.~23244057, 16H03988 for J.N.) from 
Japan Society for the Promotion of Science. 
S.S.\ was supported by the MEXT-Supported Program 
for the Strategic Research Foundation at Private Universities 
``Topological Science'' (Grant No.~S1511006).

\end{document}